\documentclass[a4paper,conference,onecolumn]{IEEEtran}

\usepackage[utf8]{inputenc}
\usepackage[T1]{fontenc}
\usepackage{url}
\usepackage{ifthen}
\usepackage{amsfonts}
\usepackage{cite}
\usepackage{tikz}
\usepackage{cancel}
\usepackage{xcolor}
\usepackage{float}
\usepackage{subcaption}
\usepackage{hyperref}
\hypersetup{pdfborder = 0 0 0}

\usepackage{amsthm}

\newtheorem{theorem}{Theorem}
\newtheorem{remark}{Remark}

\newtheorem{proposition}{Proposition}

\theoremstyle{definition}

\usepackage{amssymb}

\usetikzlibrary{patterns}
\usetikzlibrary{shapes,arrows}
\newcommand{\Exp}[2][]{\ensuremath{\mathbb{E}_{#1}\left[#2 \right]}} % $\Exp{X}{Y}$ 
\newcommand{\var}[1]{\ensuremath{{\rm var} \left[#1\right]}} % variance 
\newcommand{\cov}[1]{\ensuremath{{\rm cov} \left[#1\right]}} % covariance

\newcommand{\eqann}[2][=]{\overset{\mathclap{(\text{#2})}}{#1}} % Add (#1) over an = sign.
\newcommand{\eqannref}[1]{$(\text{#1})$}
\newcommand{\bv}[1]{\mathbf{#1}} % bold vector 
\newcommand{\rv}[1]{\mathsf{#1}} % random variable 
\newcommand{\Prob}[1]{\ensuremath{\mathbb{P} \left\{#1 \right\}}} % $\Prob{X}$ 

\newcounter{tempEquationCounter} 
\newcounter{thisEquationNumber}

\usepackage[cmex10]{mathtools} % Use the [cmex10] option 
\renewcommand{\Pr}{\mathbb{P}}

\begin{document}
\title{Parallel Gaussian Channels Corrupted by  Independent States With a State-Cognitive Helper}

%%% Single author, or several authors with same affiliation:
%\author{%
%  \IEEEauthorblockN{Michael Dikshtein}
%  \IEEEauthorblockA{Department of Electrical Engineering\\
%              ISI (D-ITET)\\
%              Haifa 32000, Israel\\
%              Email: michaeldic@campus.technion.ac.il}
%}

%%% Several authors with up to three affiliations:
% \author{%
%   \IEEEauthorblockN{Helmut Bölcskei}
%   \IEEEauthorblockA{ETH Zürich\\
%              IKT (D-ITET), ETH Zentrum\\
%              CH-8092 Zürich, Switzerland\\
%              Email: boelcskei@nari.ee.ethz.ch}
%   \and
%   \IEEEauthorblockN{Amos Lapidoth and Stefan M.~Moser}
%   \IEEEauthorblockA{ETH Zürich\\
%              ISI (D-ITET), ETH Zentrum\\
%              CH-8092 Zürich, Switzerland\\
%              Email: \{lapidoth, moser\}@isi.ee.ethz.ch}
% }

%% Many authors with many affiliations:
 \author{%
   \IEEEauthorblockN{Michael Dikshtein\IEEEauthorrefmark{1} \href{https://orcid.org/0000-0002-0498-273X}{\includegraphics[width=0.32cm]{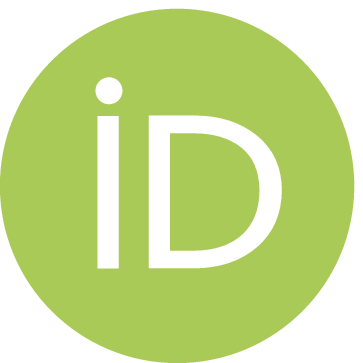}},
     Ruchen Duan\IEEEauthorrefmark{2},
     Yingbin Liang\IEEEauthorrefmark{2}\IEEEauthorrefmark{3},
     and Shlomo Shamai (Shitz)\IEEEauthorrefmark{4}}
   \IEEEauthorblockA{\IEEEauthorrefmark{1}%
              Department of EE, 
              Technion, 
              Haifa 32000, Israel,
              michaeldic@campus.technion.ac.il}
   \IEEEauthorblockA{\IEEEauthorrefmark{2}%
              Samsung Semiconductor Inc, 
              San Diego, CA 92121 USA,
              r.duan@samsung.com}
   \IEEEauthorblockA{\IEEEauthorrefmark{3}%
              Department of ECE,
              The Ohio State University,
              Columbus, OH 43210 USA, 
              liang.889@osu.edu}
   \IEEEauthorblockA{\IEEEauthorrefmark{4}%
              Department of EE, 
              Technion, 
              Haifa 32000, Israel,
              sshlomo@ee.technion.ac.il}
 }

\maketitle

\begin{abstract}
We consider a state-dependent parallel Gaussian channel with independent states and a common cognitive helper, in which two transmitters wish to send independent information to their corresponding receivers over two parallel subchannels. Each channel is corrupted by independent additive Gaussian state. The states are not known to the transmitters nor to the receivers, but known to a helper in a noncausal manner. The helper's goal is to assist a reliable communication by mitigating the state. Outer and inner bounds are derived and segments of the capacity region is characterized for various channel parameters.
\end{abstract}


\section{Introduction}
In this paper we consider a communication scenario where two transmitters wish to send messages to their corresponding receivers over a parallel state-dependent channel and a helper who knows the state in a noncausal manner, wishes to assist each receiver to mitigate the interference caused by the state. The motivation to study such a model arises from practical considerations. For example, consider a situation where there are two Device to Device (D2D) links located in two distinct cells and there is a downlink signal sent from the base-station to some conventional mobile user in the cell. In addition there is some central unit that knows in a noncausal manner the signal to be sent by each base-station, the helper in our model, and tries to assist the D2D communication links by mitigating the interference.

The model addressed in this paper has a mismatched property, that is the state sequence is known only to some nodes, which differs it from the classical study on state-dependent channels. The study of channels with side information goes back to Shannon who  considered a DMC channel with random parameters with casual side information at the transmitter. The case of noncausal side information was solved by Gel'fand and Pinsker (GP) \cite{Gelf80} for the discrete memoryless channel. Costa \cite{Costa83} considered a Gaussian version of the GP channel, and derived a surprising result, such that the interference can be completely canceled. Such a phenomena is known as Writing on Dirty Paper (WDP) property. Steinberg and Shamai \cite{shamai2005} proposed an achievable scheme for the broadcast channel with random parameters, where they have shown that the WDP property holds for the Gaussian BC with additive state. In this work a similar scheme would be used to derive an inner bound.

The type of channels with mismatched property has been addressed in the past for various models, for example, in \cite{zaidi2009} the state dependent MAC channel is studied with the state known at only one transmitter. The best outer bound for the Gaussian MAC setting was recently reported in \cite{yang2018state}. The point to point helper channel studied in \cite{malik2008} and \cite{sun2016capacity} can be considered as a special case of \cite{zaidi2009}, where the cognitive transmitter does not send any message. Authors of \cite{boostanpour2018impact} have recently considered a scenario with a state cognitive relay. The state dependent Z-IC with common state known in noncausal manner only to the primary user was studied in \cite{ghasemi2018capacity}.

Our previous work \cite{dikshtein2018state} studied a situation where each channel is corrupted by same but differently scaled state was considered. In \cite{duan2015} a similar setup was considered but with infinite state power. The achievabilty scheme in latter paper was a time-sharing version of point-to-point helper channel, such that the helper alternatively assists receivers. This work differs from the previous ones in that we address a situation where the states are independent with arbitrary state power.

Our main contribution in this paper is derivation of inner bound which is an extension of the Marton coding scheme for discrete broadcast channel to the current model. We will apply this bound for the Gaussian setting and characterize the segments of the capacity region for various channel parameters.

\section{Notations and Problem Formulation}

Random variables are denoted using a sans-serif font, e.g., $ \rv{X} $, their realizations are denoted by the respective lower case letters, e.g., $ x $, and their alphabets are denoted by the respective calligraphic letter, e.g., $ \mathcal{X} $. The expectation of $ \rv{X} $ is denoted by $ \Exp{\rv{X}} $. Let $ \mathcal{X}^n $ stand for the set of all $ n $-tuples of elements from $ \mathcal{X} $. An element from $ \mathcal{X}^n $ is denoted by $ x^n = (x_1,x_2,\dots , x_n) $ and substrings by $ x_{i}^j = (x_i,x_{i+1},\dots ,x_j) $.

\begin{figure}
	\centering
	
	\tikzstyle{block} = [draw, rectangle, minimum width = 1cm, minimum height = 0.75cm]
	\tikzstyle{sum} = [draw,circle]
	\tikzstyle{input} = [coordinate]
	\tikzstyle{dummy} = [coordinate]
	\tikzstyle{output} = [coordinate]
	\tikzstyle{amp} = [draw,shape border rotate = 180,regular polygon,regular polygon sides=3]
	
	\begin{tikzpicture}
		% Helper nodes
		\node [input]	(in0)	at	(0,-1.5)	{};
		\node [block]	(enc0)	at	(1,-1.5)	{Helper};	
		
		% Encoder 1 nodes
		\node [input]	(in1)	at 	(0,0) 		{};
		\node [block]	(enc1) 	at 	(1,0)	{Enc $1$};
		\node [sum]		(sum11)	at 	(2.75,0)		{$+$};
		\node [sum]		(sum12)	at 	(4,0)		{$+$};
		\node [sum]		(sum13)	at	(5.25,0)		{$+$};
		\node [block]	(dec1)	at	(6.75,0)		{Dec $ 1 $};
		\node [output]	(out1)	at	(7.75,0)		{};
		
		% State node
		% Encoder 2 nodes
		\node [input]	(in2)	at	(0,-3)		{};
		\node [block]	(enc2)	at	(1,-3)		{Enc 2};
		\node [sum]		(sum21)	at	(2.75,-3)		{$+$};
		\node [sum]		(sum22)	at	(4,-3)		{$+$};
		\node [sum]		(sum23)	at	(5.25,-3)		{$+$};
		\node [block]	(dec2)	at	(6.75,-3)		{Dec 2};
		\node [output]	(out2)	at	(7.75,-3)		{};
		\draw [->] (enc0) -- ++(1,0) node[above] {$\rv{X}_0^n$} --  (enc0-|sum11)--node[right]{$ \eta_1 $} (sum11) ;
		\draw [->] (sum13) ++(0,-1) node[below, anchor=west] {$\rv{Z}_1^n\sim \mathcal{N}(0,1)$}--   (sum13);	
		\draw [->] (in1) node[left] {$\rv{M}_1$} --  (enc1);
		\draw [->] (enc1) -- ++(1,0)node[above] {$\rv{X}_1^n$} -- (sum11);
		\draw [->] (sum11) -- (sum12);
		\draw [->] (sum12) -- (sum13);
		\draw [->] (sum13) -- node[above] {$\rv{Y}^n_1$}  (dec1);
		\draw [->] (dec1) --  (out1) node[right] {$\hat{\rv{M}}_1$};
		\draw [->] (in0) node[left] {$S^n$} -- node[pos=0.1] {} (enc0);
		\draw [->] (sum23) ++(0,1)node[above, anchor=west] {$\rv{Z}_2^n\sim \mathcal{N}(0,1)$} --  (sum23);
		\draw [->] (in2) node[left] {$\rv{M}_2$} -- (enc2);
		\draw [->] (enc2) -- ++(1,0) node[above] {$\rv{X}_2^n$}  -- (sum21);
		\draw [->] (sum21) -- (sum22);
		\draw [->] (sum22) -- (sum23);
		\draw [->] (sum23) -- node[above] {$\rv{Y}_2^n$} (dec2);
		\draw [->] (dec2) -- (out2)	node[right] {$\hat{\rv{M}}_2$} ;
		\fill (enc0-|sum11) circle [radius = 2pt];	
		\draw [->] (sum12) ++(0,-0.75) node[below] {$ \rv{S}^n_1 $}--   (sum12);
		\draw [->] (sum22) ++(0,0.75) node[above] {$\rv{S}_2^n$}--   (sum22);
		\draw [->] (enc0-|sum21)-- node[right] {$ \eta_2 $} (sum21);
	\end{tikzpicture}
	\caption{State-Dependent Parallel Channel with a Helper.}
	\label{fig:channel_model}
\end{figure}
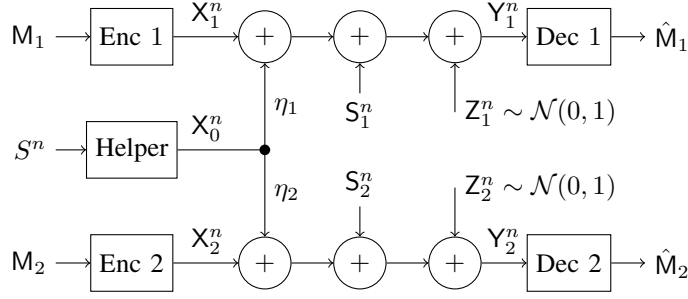

We consider a 3-transmitter, 2-receiver \textit{state dependent parallel discrete memoryless channel} depicted in Figure \ref{fig:channel_model}, where Transmitter 1 wishes to communicate a message $ \rv{M}_1 $ to Receiver 1, and similarly Transmitter 2 wishes to transmit a message $ \rv{M}_2 $ to its corresponding Receiver 2. The messages $ \rv{M}_1 $ and $ \rv{M}_2 $ are independent. The communication takes over a parallel state-dependent channel characterized by a probability transition matrix $ p(y_1,y_2|x_0,x_1,x_2,s) $. The Transmitter at the helper has noncausal knowledge of the state and tries to mitigate the interference caused in both channels.   The state variable $ \rv{S} $ is random taking values in $ \mathcal{S} $ and drawn from a discrete memoryless source (DMS)
\begin{align*}
	P_{\rv{S}^n} (s^n) = \prod_{i=1}^{n} P_\rv{S}(s_i)
\end{align*}
A $(2^{nR_1},2^{nR_2},n) $ code for the parallel state-dependent channel with state known non-causally at the helper consists of
\begin{itemize}
	\item Two message sets $ [1:2^{nR_1} \} $ and $ [1:2^{nR_2}] $.
	\item Three encoders, where encoder at the helper assigns a sequence $ x_0^n(s^n) $ to each state sequence $ s^n \in \mathcal{S}^n $, encoder 1 assigns a codeword $ x_1^n(m_1) $ to each message $ m_1 \in [1:2^{nR_1}] $ and encoder 2 assigns a codeword $ x_2^n(m_2) $ to each message $ m_2 \in [1:2^{nR_2}] $.
	\item Two decoders, where decoder 1 assigns an estimate $ \hat{m}_1 \in [1:2^{nR_1}] $ or an  error message e to each received sequence $ y_1^n $, and decoder 2 assigns an estimate $ \hat{m_2} \in [1:2^{nR_2}] $ or an error message e to each received sequence $ y_2^n $.
\end{itemize}
We assume that the message pair $ (\rv{M}_1, \rv{M}_2) $ is uniformly distributed over $ [1:2^{nR_1}] \times [1:2^{nR_2}] $. The average probability of error for a length-$ n $ code is defined as
\begin{equation}
	P_e^{(n)} = \Prob{\hat{\rv{M}}_1 \neq \rv{M}_1 \text{ or } \hat{\rv{M}}_2 \neq \rv{M}_2}.
\end{equation}
A rate pair $ (R_1,R_2) $ is said to be achievable if there exists a sequence of $ (2^{nR_1},2^{nR_2},n) $ codes such that $ \lim_{n\rightarrow \infty} P_e^{(n)} = 0$. The capacity region $ \mathcal{C}$ is the closure of the set of all achievable rate pairs $ (R_1,R_2) $.

Our goal is to characterize the capacity region $ \mathcal{C} $ for the state-dependent Gaussian parallel channel with additive state known at the helper, where the outputs at the receivers for one channel use are described by the equations
	\begin{subequations}
		\begin{equation}
			\rv{Y_1} = \eta_1 \rv{X_0}+\rv{X_1}+\rv{S_1}+\rv{Z_1}
		\end{equation}
		\begin{equation}
			\rv{Y_2} = \eta_2 \rv{X_0}+\rv{X_2}+\rv{S_2}+\rv{Z_2}
		\end{equation}
	\end{subequations}
	where $ \rv{Z_1} \sim \mathcal{N} (0,1) $ and $ \rv{Z_2} \sim \mathcal{N} (0,1) $ are additive Gaussian noise of $ \rv{Y_1} $ and $ \rv{Y_2} $,  $ \rv{S_1} \sim \mathcal{N} (0,Q_1) $ and $ \rv{S_2} \sim \mathcal{N} (0,Q_2) $ are  additive Gaussian state, both known noncausally at the transmitter, and $ \eta_j $, $ j=1,2 $, is the channel gain from the helper to receiver $ j $. The Gaussian random variables $ \rv{Z_1} $, $ \rv{Z_2} $, $ \rv{S_1} $, $ \rv{S_2} $ are independent of each other. The channel inputs $ \rv{X_j} $, $ j=0,1,2 $ are power constrained: $ \Exp{\rv{X_j}^2} \leq P_j $. 

\section{Main Results}\label{sec:result}

\subsection{Outer and Inner Bounds} \label{subsection_independent_states_bounds}
	In order to characterize the capacity region of this channel, we first provide an outer bound on the capacity region as follows
	\begin{proposition} \label{proposition:independent_states_outer_bound}
		Every achievable rate pair $ (R_1,R_2) $ of the state-dependent parallel Gaussian channel with a helper must satisfy the following inequalities
		\begin{subequations}
			\begin{equation} \label{eq:independent_states_R1_upper_bound}
				R_1 \leq \min \bigg\{ \frac{1}{2} \log \left( 1+\frac{P_1}{\eta_1^2 P_0+2 \eta_1 \rho_{0S_1} \sqrt{P_0 Q_1}+Q_1+1} \right) \\
				+\frac{1}{2} \log \left( (1-\rho_{0S_1}^2-\rho_{0S_2}^2)\eta_1^2P_0+1 \right), \frac{1}{2} \log (1+P_1) \bigg\}
			\end{equation}
			\begin{equation} \label{eq:independent_states_R2_upper_bound}
				R_2 \leq \min \bigg\{ \frac{1}{2} \log \left( 1+\frac{P_2}{\eta_2^2 P_0+2 \eta_2 \rho_{0S_2} \sqrt{P_0 Q_2}+Q_2+1} \right) \\
				+\frac{1}{2} \log \left( (1-\rho_{0S_1}^2-\rho_{0S_2}^2)\eta_2^2P_0+1 \right), \frac{1}{2} \log (1+P_2) \bigg\}
			\end{equation}
			for some $ \rho_{0S_1} $ and $ \rho_{0S_2} $ that satisfy 
			\begin{equation} \label{eq:independent_states_outer_bound_rho_constraint}
				\rho_{0S_1}^2+\rho_{0S_2}^2 \leq 1.
			\end{equation}
		\end{subequations}
	\end{proposition}
	\begin{IEEEproof}
		This outer bound is an extension of the outer bound derived in \cite{malik2008}. For a complete proof see Appendix \ref{apndx:independent_states_outer_bound}.
	\end{IEEEproof}
The upper bound for each rate consists of two terms, the first one reflects the scenario when the interference cannot be completely canceled, and the second is simply the point-to-point capacity of channel without state.

We next derive an achievable region for the channel based on an achievable scheme that integrates Marton's coding, single-bin dirty paper coding and state cancellation. More specifically, we generate two auxiliary random variables, $ \rv{U} $ and $ \rv{V} $ to incorporate the state information so that Receiver 1 (and respectively 2) decodes $ \rv{U} $ (and respectively $ \rv{V} $) and then decodes the respective transmitter information. Based on such achievable scheme, we derive the following inner bound on the capacity region for the DM case.

\begin{proposition} \label{prp:dm_independent_states_inner_bound}
	An inner bound on the capacity region of the discrete memoryless parallel state-dependent channel with a helper consists of rate pairs $ (R_1,R_2) $ satisfying:
	\begin{subequations}
		\begin{equation}
			R_1 \leq \min \{I(\rv{U},\rv{X_1};\rv{Y_1})-I(\rv{U};\rv{S}),I(\rv{X_1};\rv{Y_1}|\rv{U})\} 
		\end{equation}
		\begin{equation}
			R_2 \leq \min \{I(\rv{V},\rv{X_2};\rv{Y_2})-I(\rv{V};\rv{S}),I(\rv{X_2};\rv{Y_2}|\rv{V})\} 
		\end{equation}
		\begin{equation}
			R_1+R_2 \leq \min \{I(\rv{U},\rv{X_1};\rv{Y_1})-I(\rv{U};\rv{S})+I(\rv{V},\rv{X_2};\rv{Y_2})
			-I(\rv{V};\rv{S})-I(\rv{V};\rv{U}|\rv{S}),
			I(\rv{X_1};\rv{Y_1}|\rv{U})+I(\rv{X_2};\rv{Y_2}|\rv{V})\}
		\end{equation}
%		\begin{multline}
%			R_1+R_2 \leq \min \{I(\rv{U},\rv{X_1};\rv{Y_1})-I(\rv{U};\rv{S})+I(\rv{V},\rv{X_2};\rv{Y_2})\\
%			\quad -I(\rv{V};\rv{S})-I(\rv{V};\rv{U}|\rv{S}),\\
%			\quad I(\rv{X_1};\rv{Y_1}|\rv{U})+I(\rv{X_2};\rv{Y_2}|\rv{V})\}
%		\end{multline}
	\end{subequations}
	for some pmf $ p(u,v,x_0|s)p(x_1)p(x_2) $.
\end{proposition}
\begin{remark} \label{remark:dm_independent_states_inner_bound}
	The achievable region in Proposition \ref{prp:dm_independent_states_inner_bound} is equivalent to the following region
	\begin{subequations}
		\begin{equation}
		R_1 \leq \min \{I(\rv{U}_1,\rv{X_1};\rv{Y_1})-I(\rv{U}_1;\rv{S}),I(\rv{X_1};\rv{Y_1}|\rv{U}_1)\} 
		\end{equation}
		\begin{equation}
		R_2 \leq \min \{I(\rv{U}_2,\rv{X_2};\rv{Y_2})-I(\rv{U}_2;\rv{U}_1,\rv{S}),I(\rv{X_2};\rv{Y_2}|\rv{U}_2)\}
		\end{equation}
		for some pmf $ p(u_1,u_2,x_0|s)p(x_1)p(x_2) $.
	\end{subequations}
\end{remark}
\begin{IEEEproof}
	See Appendix \ref{apndx:independent_state_inner_bound}.
\end{IEEEproof}

Denote 
\begin{align*}
\overline{\alpha}_1 \triangleq (\alpha_{11},\alpha_{12}) \quad	\overline{\alpha}_2 \triangleq (\alpha_{20},\alpha_{21},\alpha_{22}) \quad 
\overline{\beta} \triangleq (\beta_1,\beta_2).
\end{align*}
Let $ f_1(\cdot) ,g_1(\cdot), f_2(\cdot)$ and $ g_2(\cdot) $ be defined as
\begin{equation*}
f_k (\overline{\alpha}_k,\overline{\beta},\gamma) \!=\!	\frac{1}{2} \log \frac{\eta_k^2 \gamma P'_{0} \cdot \sigma^2_{Y_k} (\beta_k) } {h_k (\overline{\alpha}_k,\overline{\beta},\gamma) } 
\end{equation*}
\begin{equation*}
g_k (\overline{\alpha}_k,\overline{\beta},\gamma)
= \frac{1}{2} \log \left( 1+\frac{P_{k} \cdot\sigma^2_{U_k} (\overline{\alpha}_{k}) } {h_k (\overline{\alpha}_k,\overline{\beta},\gamma)} \right)
\end{equation*}
where
\begin{equation*}
h_k (\overline{\alpha}_k,\overline{\beta},\gamma) =	\sigma^2_{Y_k|X_k} (\beta_k) \cdot \sigma^2_{{U}_k} (\overline{\alpha}_{k})-\sigma_{U_kY_k}^2 (\overline{\alpha}_{k},\overline{\beta})  
\end{equation*}
\begin{equation*}
\sigma^2_{Y_k} (\beta_k) = \eta_k^2 P_0+(2\beta_k \eta_k +1)Q_k+P_k+1
\end{equation*}
\begin{equation*}
\sigma^2_{Y_k|X_k} (\beta_k) = \eta_k^2 P_0+(2\beta_k \eta_k +1)Q_k+1
\end{equation*}
\begin{equation*}
\sigma^2_{U_1} (\overline{\alpha}_{1}) = \eta_1^2 \gamma P'_{0}+\alpha_{11}^2Q_1+\alpha_{12}^2 Q_2
\end{equation*}
\begin{equation*}
\sigma^2_{U_2} (\overline{\alpha}_{2}) = \eta_2^2(\overline{\gamma}+\alpha_{20}^2 \gamma) P'_{0}+\alpha_{21}^2Q_1+\alpha_{22}^2 Q_2
\end{equation*}
\begin{equation*}
\sigma_{U_1,Y_1} (\overline{\alpha}_{1},\overline{\beta})= \eta_1^2 \gamma P'_{0}+(1+\beta_1\eta_1)\alpha_{11} Q_1+\alpha_{12}\beta_2 \eta_1 Q_2
\end{equation*}
\begin{equation*}
\sigma_{U_2,Y_2} (\overline{\alpha}_{2},\overline{\beta}) \!=\! \eta_2^2 (P'_{02}+\alpha_{20} P'_{01})+\alpha_{22} Q_2(1+\beta_2 \eta_2)+\alpha_{21}\beta_1\eta_2Q_1
\end{equation*}
 Based on the above inner bound, we obtain an achievable region for the Gaussian channel by setting an appropriate joint input distribution.

\begin{proposition} \label{prp:Gaussian_independent_states_inner_bound}
	An inner bound on the capacity region of the parallel state-dependent Gaussian channel with a helper consists of rate pairs $ (R_1,R_2) $ satisfying;
	\begin{subequations}
		\begin{equation} \label{eq:independent_states_R1_inner_bound1}
			R_1 \leq \min \{ f_1 (\overline{\alpha}_1,\overline{\beta},\gamma) ,g_1 (\overline{\alpha}_1,\overline{\beta},\gamma) \}
		\end{equation}
		\begin{equation} \label{eq:independent_states_R2_inner_bound1}
		R_2 \leq \min \{  f_2 (\overline{\alpha}_2,\overline{\beta},\gamma),g_2 (\overline{\alpha}_1,\overline{\beta},\gamma) \}
		\end{equation}
	\end{subequations}
	for some real constants $ \alpha_{11} $, $ \alpha_{12} $, $ \alpha_{20} $, $ \alpha_{21} $, $ \alpha_{22} $, $ \beta_1 $, $ \beta_2 $ and $ \gamma $ satisfying $ \beta_1^2 Q_1+	\beta_2^2 Q_2 \leq P_0 $, $ \gamma \in [0,1] $ and $ \overline{\gamma}=1-\gamma $.
\end{proposition}
\begin{IEEEproof}
	The region follows from Remark \ref{remark:dm_independent_states_inner_bound} by choosing the joint Gaussian distribution for random variables as follows:
	\begin{align*}
		& \rv{U} = \rv{X'_{01}}+\eta_1^{-1}(\alpha_{11} \rv{S_1}+\alpha_{12} \rv{S_2}) \\
		& \rv{V} = \rv{X'_{02}}+\alpha_{20} \rv{X'_{01}}+\eta_2^{-1} (\alpha_{21}\rv{S_1}+\alpha_{22} \rv{S_2}) \\
		& \rv{X_0} = \rv{X'_{01}}+\beta_1 \rv{S_1}+\rv{X'_{02}}+\beta_2 \rv{S_2} \\
		& \rv{X'_{01}} \sim \mathcal{N} (0,\gamma P'_{0}) \quad  \rv{X'_{02}} \sim \mathcal{N} (0, \overline{\gamma} P'_{0}) \\
		& \rv{X_1} \sim \mathcal{N} (0,P_{1}) \quad \rv{X_2} \sim \mathcal{N} (0,P_{2}) 
	\end{align*}
	where $ \rv{X'_{01}}, \rv{X'_{02}},\rv{X_1},\rv{X_2},\rv{S_1},\rv{S_2} $ are independent. The constraint on $ \beta_1 $ and $ \beta_2 $ follows from power constraint on $ \rv{X_0} $.
\end{IEEEproof}

Now we provide our intuition behind such construction of the r.v.'s in the proof of Proposition \ref{prp:Gaussian_independent_states_inner_bound}. $ \rv{X_0} $ contains two parts, the one with $ \beta_j $, $ j=1,2 $ controls the direct state cancellation of each state. The second part $ \rv{X'_{0j}} $, $ j=1,2 $, is used for dirty paper coding via generation of the state-correlated auxiliary r.v.'s $ \rv{U} $ and $ \rv{V} $.

	Another important result of Proposition \ref{prp:Gaussian_independent_states_inner_bound} is a constraint on $ \beta_1 $ and $ \beta_2 $
	\begin{equation} \label{eq:independent_states_beta_constraint}
		\beta_1^2 Q_1+	\beta_2^2 Q_2 \leq P_0
	\end{equation}
	we now define $ \beta_j \triangleq \rho_{0S_j} \sqrt{\frac{P_0}{Q_j}} $, and use this setting to write \eqref{eq:independent_states_beta_constraint} as
	\begin{equation}
		\rho_{0S_1}^2+\rho_{0S_2}^2 \leq 1
	\end{equation}
	which is equivalent to \eqref{eq:independent_states_outer_bound_rho_constraint}.
\subsection{Capacity Region Characterization}
In this section we will characterize segments on the capacity boundary for various channel parameters using the inner and outer bounds that were derived in Section \ref{subsection_independent_states_bounds}. Consider the inner bounds in \eqref{eq:independent_states_R1_inner_bound1} - \eqref{eq:independent_states_R2_inner_bound1}. Each bound has two terms in the argument of min. We suggest to optimize each term independently and then compare it to the outer bounds in \eqref{eq:independent_states_R1_upper_bound}-\eqref{eq:independent_states_R2_upper_bound}. In the last step we will state the conditions under which those terms are valid.
We first consider the bound on $ R_1 $. Let
\begin{align}
	\alpha_{11}^a = \frac{(1+\eta_1 \beta_1)\eta_1^2 \gamma P'_{0}}{\eta_1^2 P'_{0}+1}\quad
	\alpha_{12}^a = \frac{\beta_2 \eta_1^3  \gamma P'_{0}}{\eta_1^2 P'_{0}+1}
\end{align}
Then $ f_1(\overline{\alpha}_1^a,\overline{\beta},\gamma) $ takes the following form
\begin{equation} \label{eq:inner_bound_f1}
f_1(\overline{\alpha}_1^a,\overline{\beta},\gamma)=  \frac{1}{2} \log \left( 1+\frac{P_1}{\eta_1^2 P_0+2\eta_1 \rho_{0S_1} \sqrt{P_0 Q_1}+Q_1+1} \right)
+ \frac{1}{2} \log \left(1+ \frac{\eta_1^2 \gamma  P'_{0}}{1+\eta_1^2 \overline{\gamma}P'_{0}} \right)
\end{equation}
%\begin{equation} \label{eq:inner_bound_f1}
%\begin{split}
%&f_1(\overline{\alpha}_1^a,\overline{\beta},\gamma)\\ &\quad=  \frac{1}{2} \log \left( 1+\frac{P_1}{\eta_1^2 P_0+2\eta_1 \rho_{0S_1} \sqrt{P_0 Q_1}+Q_1+1} \right) \\
%&\qquad+ \frac{1}{2} \log \left(1+ \frac{\eta_1^2 \gamma  P'_{0}}{1+\eta_1^2 \overline{\gamma}P'_{0}} \right)
%\end{split}
%\end{equation}
If $ f_1(\overline{\alpha}_1^a,\overline{\beta},\gamma) \leq g_1(\overline{\alpha}_1^a,\overline{\beta},\gamma) $, then $ R_1 = f_1(\overline{\alpha}_1^a,\overline{\beta},\gamma) $ is achievable. 
Moreover, if we choose $ \gamma=1 $, then $ R_1 =  f_1(\overline{\alpha}_1^a,\overline{\beta},1)$ meets the outer bound (the first term in "min" in \eqref{eq:independent_states_R1_upper_bound}).
Furthermore, by setting
\begin{align*}
\alpha^{b}_{11}=1+\eta_1 \beta_1 \qquad \alpha^{b}_{12}=\eta_1 \beta_2
\end{align*}
we obtain
\begin{align*}
g_1(\overline{\alpha}_1^b,\overline{\beta},\gamma) = \frac{1}{2} \log \left(1+\frac{P_1}{1+\eta_1^2 \overline{\gamma} P'_{0}} \right)
\end{align*}
If $ g_1(\overline{\alpha}_1^b,\overline{\beta},\gamma) \leq f_1(\overline{\alpha}_1^b,\overline{\beta},\gamma) $, then 
\begin{align*}
	R_1 = \frac{1}{2} \log \left(1+\frac{P_1}{1+\eta_1^2 \overline{\gamma} P'_{0}} \right) 
\end{align*}
is achievable. 
Similarly, by choosing $ \gamma=1 $, then $ R_1 = \frac{1}{2} \log (1+P_1)$ is achievable and this meets the outer bound (the second term in "min" in \eqref{eq:independent_states_R1_upper_bound}).
Next we consider the bound on $ R_2 $. Let
\begin{equation}
		\alpha_{20}^a=\frac{ \eta_2^2 \overline{\gamma} P'_{0}}{\eta_2^2 \overline{\gamma} P'_{0}+1}  \qquad
		\alpha_{21}^a = \frac{\beta_1 \eta_2^3  \overline{\gamma} P'_{0}}{\eta_2^2\overline{\gamma} P'_{0}+1} \qquad	 \alpha_{22}^a = \frac{(1+\eta_2 \beta_2)\eta_2^2 \overline{\gamma} P'_{0}}{\eta_2^2\overline{\gamma} P'_{0}+1}
\end{equation}
%\begin{equation}
%\begin{split}
%\alpha_{20}^a=\frac{ \eta_2^2 \overline{\gamma} P'_{0}}{\eta_2^2 \overline{\gamma} P'_{0}+1}  \quad
%\alpha_{21}^a = \frac{\beta_1 \eta_2^3  \overline{\gamma} P'_{0}}{\eta_2^2\overline{\gamma} P'_{0}+1} \\
%\quad \alpha_{22}^a = \frac{(1+\eta_2 \beta_2)\eta_2^2 \overline{\gamma} P'_{0}}{\eta_2^2\overline{\gamma} P'_{0}+1}
%\end{split}
%\end{equation}
Then $ f_2(\overline{\alpha}_2,\overline{\beta},\gamma) $ takes the following form
\begin{equation} \label{eq:inner_bound_f2}
		f_2(\overline{\alpha}_2^a,\overline{\beta},\gamma) 
		= \frac{1}{2} \log \left( 1+\frac{P_2}{\eta_2^2 P_0+2\eta_2 \rho_{0S_2} \sqrt{P_0 Q_2}+Q_2+1} \right) 
		+ \frac{1}{2} \log \left( 1+\eta_2^2 \overline{\gamma} P'_{0} \right)
\end{equation}
%\begin{equation} \label{eq:inner_bound_f2}
%\begin{split}
%&f_2(\overline{\alpha}_2^a,\overline{\beta},\gamma) \\
%&= \frac{1}{2} \log \left( 1+\frac{P_2}{\eta_2^2 P_0+2\eta_2 \rho_{0S_2} \sqrt{P_0 Q_2}+Q_2+1} \right) \\
%&+ \frac{1}{2} \log \left( 1+\eta_2^2 \overline{\gamma} P'_{0} \right)
%\end{split}
%\end{equation}
If $ f_2(\overline{\alpha}_2^a,\overline{\beta},\gamma) \leq g_2(\overline{\alpha}_2^a,\overline{\beta},\gamma) $, then $ R_2 = f_2(\overline{\alpha}_2^a,\overline{\beta},\gamma) $ is achievable. Moreover, if we choose $ \gamma=0 $, then $ R_2 =  f_2(\overline{\alpha}_2^a,\overline{\beta},0)$ meets the outer bound (the first term in "min" in \eqref{eq:independent_states_R2_upper_bound}).

Furthermore, we set 
\begin{equation}
	\alpha^{b}_{20}=1 \qquad \alpha^{b}_{21}=\eta_2 \beta_1 \qquad \alpha^{b}_{22}=1+\eta_2 \beta_2
\end{equation}
and then obtain
\begin{equation}
	g_2(\overline{\alpha}_2^b,\overline{\beta},\gamma) = \frac{1}{2} \log \left(1+P_2 \right)
\end{equation}
If $ g_2(\overline{\alpha}_2^b,\overline{\beta},\gamma) \leq f_2(\overline{\alpha}_2^b,\overline{\beta},\gamma) $, then $ R_2 = \frac{1}{2} \log \left(1+P_2 \right) $ is achievable and this meets the outer bound. This also equals the maximum rate for $ R_2 $ when the channel is not corrupted by state. 

For simplicity of representation, we denote $ \overline{\eta} \triangleq (\eta_1,\eta_2) $, $ \overline{P} \triangleq (P_1,P_2) $, $ \overline{Q} \triangleq (Q_1,Q_2) $.

Summarizing the above analysis, we obtain the following characterization of segments of the capacity region boundary.	
\begin{theorem}\label{thr:capacity}
	For every choice of $ \gamma $, the channel parameters $(\overline{\eta},P_0,\overline{P},\overline{Q})$ can be partitioned into the sets $\mathcal{A}_1,\mathcal{B}_1,\mathcal{C}_1$, where
	\begin{align*}
	\mathcal{A}_1=&\ \{(\overline{\eta},P_0,\overline{P},\overline{Q}): f_{1} (\overline{\alpha}_1^a,\overline{\beta},\gamma) \leq g_{1}(\overline{\alpha}_1^a,\overline{\beta},\gamma)\\
	\mathcal{C}_1=&\ \{(\overline{\eta},P_0,\overline{P},\overline{Q}): f_{1} (\overline{\alpha}_1^b,\overline{\beta},\gamma) \geq g_{1}(\overline{\alpha}_1^b,\overline{\beta},\gamma) \}\\
	\mathcal{B}_1 =&\ (\mathcal{A}_1 \cup \mathcal{C}_1)^c.
	\end{align*}
	If $(\overline{\eta},P_0,\overline{P},\overline{Q})\in \mathcal{A}_1$, then $ R_1 = f_{1} (\overline{\alpha}_1^a,\overline{\beta},1) $  captures one segment of the capacity region boundary, where the state cannot be fully cancelled. If $(\overline{\eta},P_0,\overline{P},\overline{Q}) \in \mathcal{C}_1$, then  $ R_1 = \frac{1}{2} \log (1+P_1) $ captures one segment of the capacity region boundary where the state is fully cancelled. If $(\overline{\eta},P_0,\overline{P},\overline{Q}) \in \mathcal{B}_1$, then the $R_1$ segment of the capacity region boundary is not characterized.
	
	The channel parameters $(\overline{\eta},P_0,\overline{P},\overline{Q})$ can also be partitioned into the sets $\mathcal{A}_2,\mathcal{B}_2,\mathcal{C}_2$, where
	\begin{align*}
	\mathcal{A}_2=&\ \{(\overline{\eta},P_0,\overline{P},\overline{Q}): f_{2} (\overline{\alpha}_2^a,\overline{\beta},\gamma) \leq g_{2}(\overline{\alpha}_2^a,\overline{\beta},\gamma)\\
	\mathcal{C}_2=&\ \{(\overline{\eta},P_0,\overline{P},\overline{Q}): f_{2} (\overline{\alpha}_2^b,\overline{\beta},\gamma) \geq g_{2}(\overline{\alpha}_2^b,\overline{\beta},\gamma) \\
	\mathcal{B}_2 =&\ (\mathcal{A}_2 \cup \mathcal{C}_2)^c.
	\end{align*}
	If $(\overline{\eta},P_0,\overline{P},\overline{Q})\in \mathcal{A}_2$, then   $ R_2 = f_{2} (\overline{\alpha}_2^a,\overline{\beta},0) $  captures one segment of the capacity region boundary, where the state cannot be fully cancelled. If $(\overline{\eta},P_0,\overline{P},\overline{Q}) \in \mathcal{C}_2$, then  $ R_2 = \frac{1}{2} \log (1+P_2)$  captures one segment of the capacity boundary where the state is fully cancelled. If $(\overline{\eta},P_0,\overline{P},\overline{Q}) \in \mathcal{B}_2$, then the $R_2$ segment of the capacity region boundary is not characterized.
	\label{thm:capacity_region}
\end{theorem}
The above theorem describes two partitions of the channel parameters, respectively under which segments on the capacity region boundary corresponding to $ R_1 $ and $ R_2 $ can be characterized. Intersection of two sets, each from one partition, collectively characterizes the entire segments on the capacity region boundary.
\subsection{Numerical Results}
\begin{figure*}
		\centering
		\begin{subfigure}[t]{0.49\textwidth}
		\centering
		\includegraphics[scale=0.6]{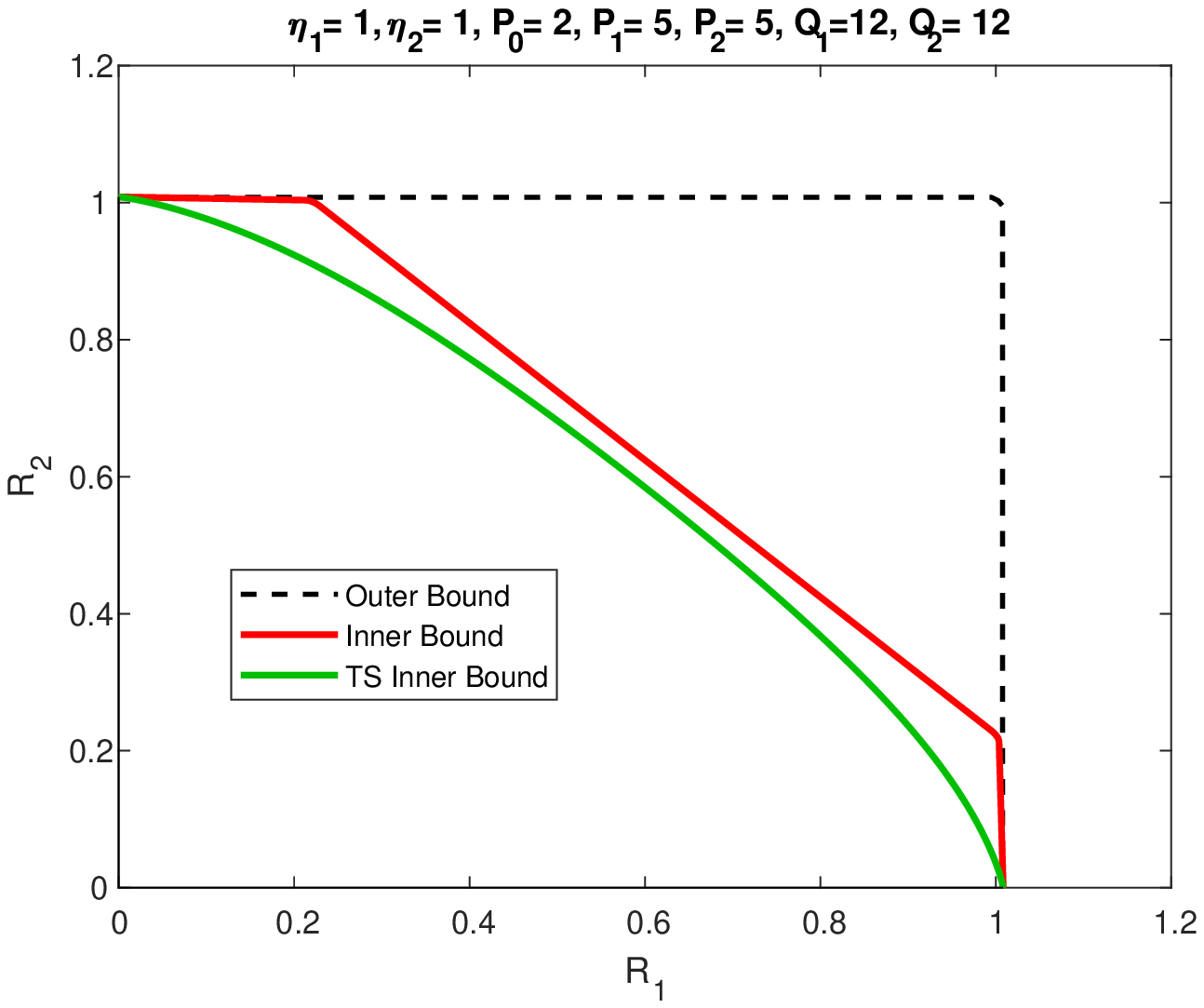}
		\caption{}
		\label{fig:independent_states_capacity_region_h1_1_h2_1_p0_2_p1_5_p2_5_q1_12_q2_12}
		\end{subfigure}
		\begin{subfigure}[t]{0.49\textwidth}
		\centering
		\includegraphics[scale=0.6]{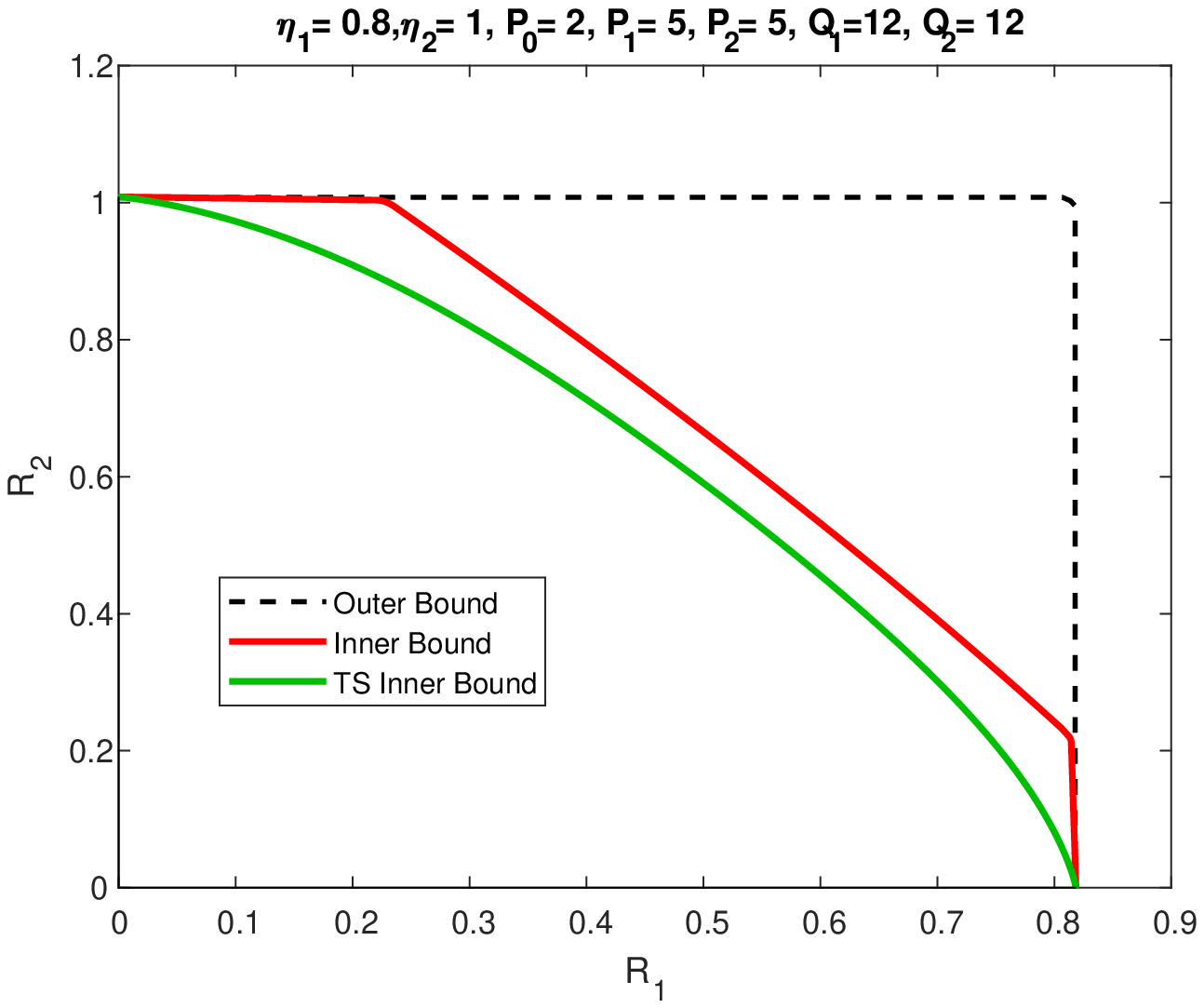}
		\caption{}
		\label{fig:independent_states_capacity_region_h1_0p8_h2_1_p0_2_p1_5_p2_5_q1_12_q2_12}
		\end{subfigure}
		\begin{subfigure}[t]{0.49\textwidth}
			\centering
			\includegraphics[scale=0.6]{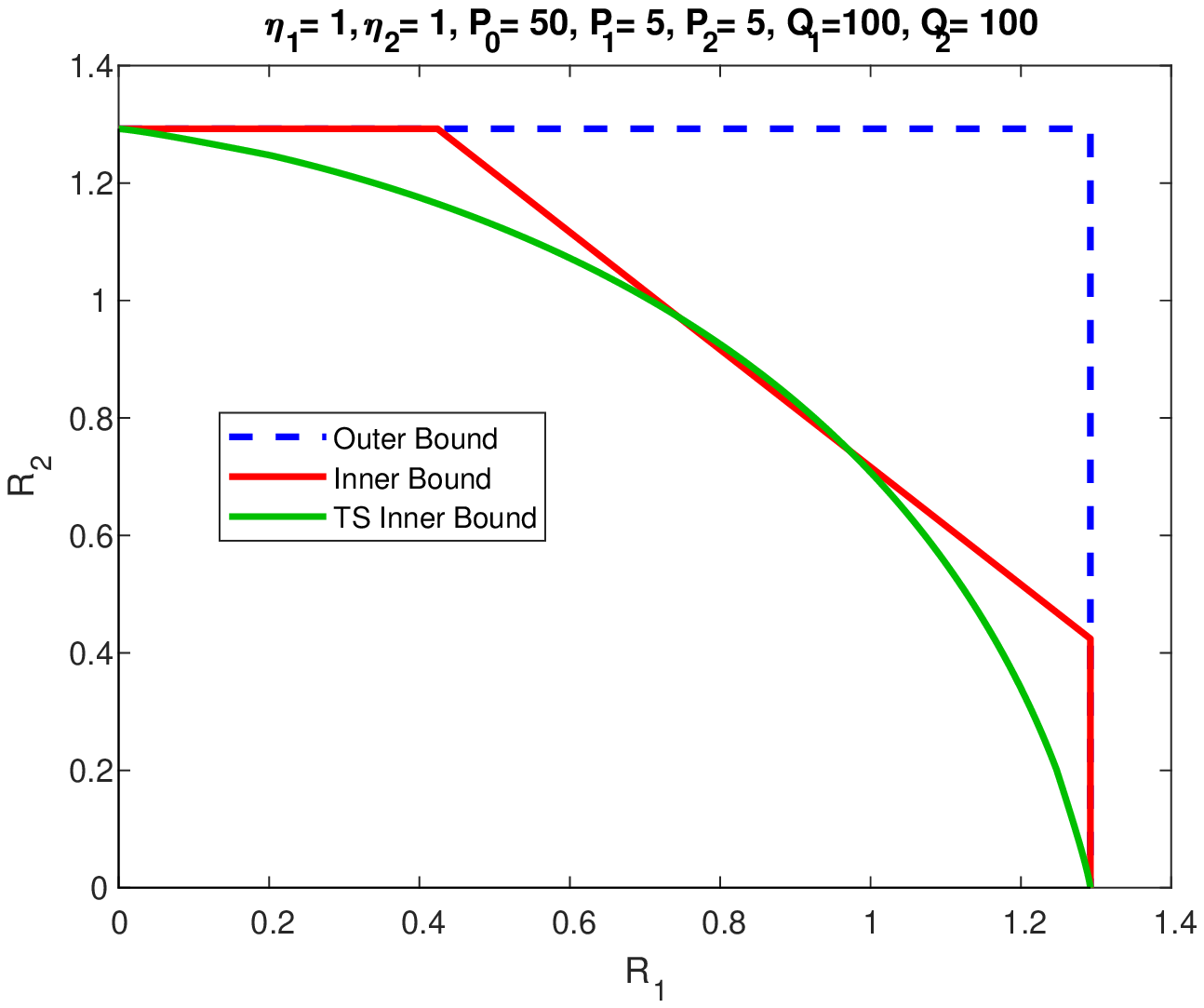}
			\caption{}
			\label{fig:independent_states_capacity_region_h1_1_h2_1_p0_50_p1_5_p2_5_q1_100_q2_100}
		\end{subfigure}
		\begin{subfigure}[t]{0.49\textwidth}
			\centering
			\includegraphics[scale=0.6]{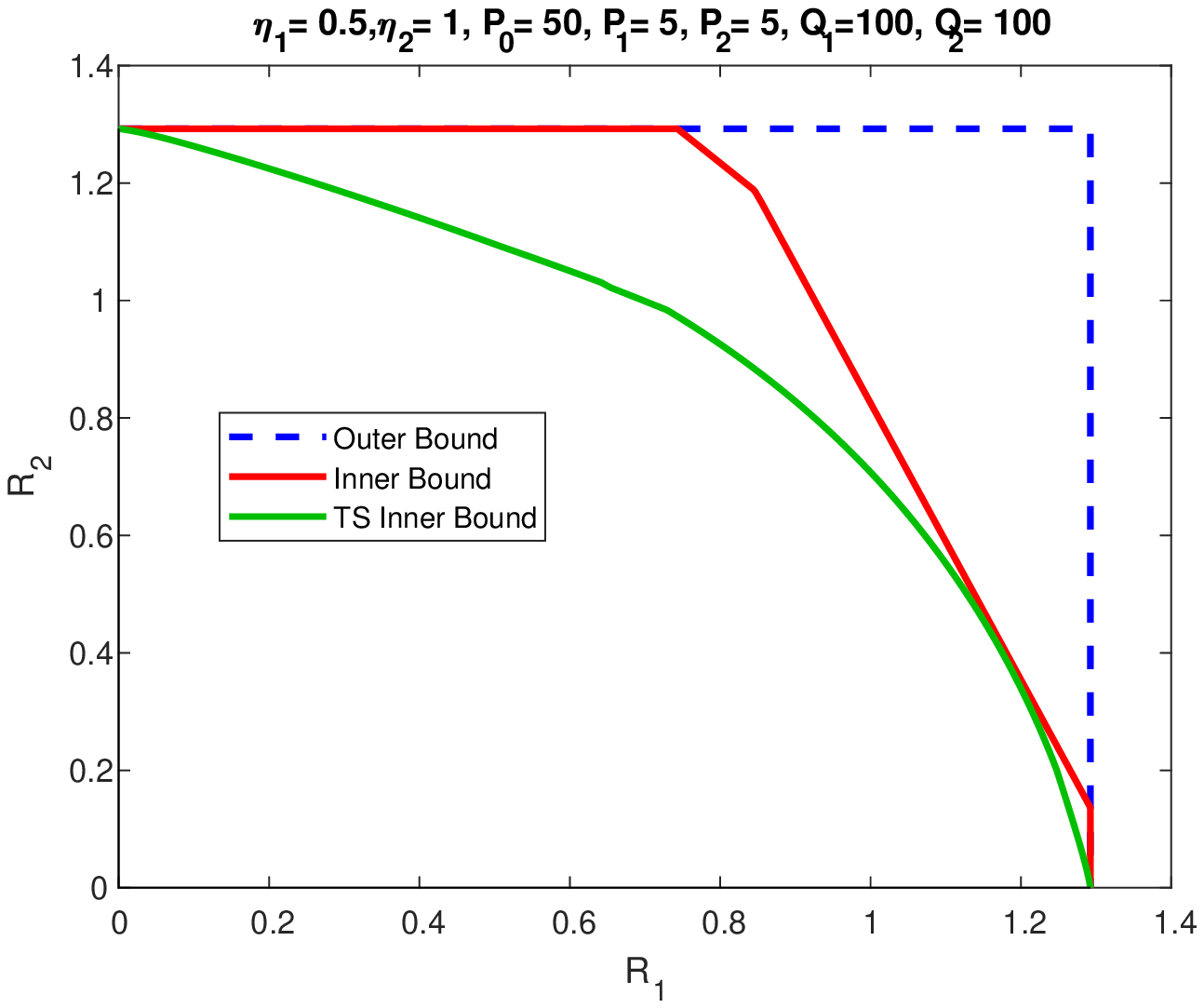}
			\caption{}
			\label{fig:independent_states_capacity_region_h1_0p5_h2_1_p0_50_p1_5_p2_5_q1_100_q2_100}
		\end{subfigure}
	\caption{Numerical Results}
	\label{fig:numerical_results}
	\end{figure*}

	In this section, we demonstrate our results using various channel parameters. We plot the inner and outer bounds for various values of helper power $ P_0 $, channel gains, $ \eta_1 $ and $ \eta_2 $ and different state power. The results are shown in Figure \ref{fig:numerical_results}. The outer bound is based on Proposition \ref{proposition:independent_states_outer_bound}. The inner bound is the convex hull of all the achievable regions, with interchange between the roles of the decoders. The time sharing inner bound is according to point-to-point helper channel achievable region. The scenario where the helper power is less than the users power is depicted in subfigures \ref{fig:independent_states_capacity_region_h1_1_h2_1_p0_2_p1_5_p2_5_q1_12_q2_12} and \ref{fig:independent_states_capacity_region_h1_0p8_h2_1_p0_2_p1_5_p2_5_q1_12_q2_12}, while the channel gains in subfigure \ref{fig:independent_states_capacity_region_h1_1_h2_1_p0_2_p1_5_p2_5_q1_12_q2_12} are equal, they are mismatched in subfigure \ref{fig:independent_states_capacity_region_h1_0p8_h2_1_p0_2_p1_5_p2_5_q1_12_q2_12}. Note that in both cases our inner bound outperforms the time-sharing bound, especially in the mismatched case, and some segments of the capacity region are characterized.
	
	The scenario with helper power being higher than the users power and matched and mismatched channel gain is depicted in subfigures  \ref{fig:independent_states_capacity_region_h1_1_h2_1_p0_50_p1_5_p2_5_q1_100_q2_100} and \ref{fig:independent_states_capacity_region_h1_0p5_h2_1_p0_50_p1_5_p2_5_q1_100_q2_100} respectively. Similarly as for low helper power regime, our proposed achievability scheme performs better than time-sharing.
	
\section{Conclusion}\label{sec:conclusion}
We have studied a  parallel state-dependent Gaussian channel with a cognitive helper with independent states and arbitrary state power. Inner and outer bounds were derived and segments of the capacity region boundary were characterized for various channel parameters. We have also demonstrated our results using numerical simulation and have shown that our achievability scheme outperforms time-sharing that was shown to be optimal for the infinite state power regime in \cite{duan2015}. In our previous work \cite{dikshtein2018state}, a model with same but differently scaled states was considered. These two models represent a special case of more general scenario with correlated states, our results in both studies imply that as the states are more correlated than it easier to mitigate the interference. Furthermore the gap between the inner bound and the outer bound in this work suggests that a new techniques for outer bound derivation is needed as we believe that the inner bounds consisting of pairs $ (R_1,R_2)=(f_1(\overline{\alpha}_1^a,\overline{\beta},\gamma), f_2(\overline{\alpha}_2,\overline{\beta},\gamma)) $ is indeed tight for some set of channel parameters.
\section*{Acknowledgment}
The work of M. Diksthein and S. Shamai has been supported by the European Union's Horizon 2020 Research And Innovation Programme, grant agreement no. 694630, and by the Heron consortium via the Israel minister of economy and science. The work of Y. Liang was supported by NSF Grant CCF-1801855.

\appendices
\section{}\label{apndx:independent_states_outer_bound}
First, define the following correlation coefficients:
\begin{equation} \label{eq:correlation_coefficients_def}
\rho_{0S_1} \triangleq \frac{\sum_{i=1}^{n} \Exp{\rv{X}_{0,i} \rv{S}_{1,i}}}{n\sqrt{P_0 Q_1 }} ,\quad \rho_{0S_2} \triangleq \frac{\sum_{i=1}^{n} \Exp{\rv{X}_{0,i} \rv{S}_{2,i}}}{n\sqrt{P_0 Q_2 }}
\end{equation}
By Fano's inequality
\begin{align}
H(\rv{M}_1|\rv{Y}_1^n) &\leq nR_1 P_e^{(n)}+1 \triangleq n \epsilon^{(1)}_n \label{eq:independent_states_fano_R1}\\
H(\rv{M}_2|\rv{Y}_2^n) &\leq nR_2P_e^{(n)}+1 \triangleq n \epsilon^{(2)}_n
\end{align}
where $ \epsilon^{(1)}_n $ and $ \epsilon^{(2)}_n $ tend to zero as $ n\rightarrow \infty $.
First we develop an upper bound on $ R_1 $ as follows:
\begin{align*}
nR_1+I(\bv{S}^n;\rv{Y}_1^n|\rv{M}_1) &= H(\rv{M}_1)-H(\rv{M}_1|\rv{Y}_1^n)+H(\rv{M}_1|\rv{Y}_1^n)+I(\bv{S}^n;\rv{Y}_1^n|\rv{M}_1) \\
&\eqann[\leq]{a} I(\rv{M}_1,\bv{S}^n;\rv{Y}_1^n)+n \epsilon^{(1)}_n\\
&= h(\rv{Y}_1^n)-h(\rv{Y}_1^n|\rv{M}_1,\bv{S}^n)+n \epsilon^{(1)}_n\\
&\eqann[\leq]{b} h(\rv{Y}_1^n)-h(\rv{Y}_1^n|\rv{M}_1,\rv{X}_1^n,\rv{X}_0^n,\bv{S}^n)+n \epsilon^{(1)}_n\\
&= h(\rv{Y}_1^n)-h(\rv{Z}_1^n)+n \epsilon^{(1)}_n \\
&\eqann[\leq]{c} \sum_{i=1}^{n} h(\rv{Y}_{1,i})-nh(\rv{Z}_1)+n \epsilon^{(1)}_n
\end{align*}
where \eqannref{a} follows from \eqref{eq:independent_states_fano_R1}, \eqannref{b} and \eqannref{c} is since conditioning reduces differential entropy. The second differential term in the last inequality is equal to $ h(\rv{Z}_1) = \frac{1}{2} \log (2\pi e) $. As for the first term in the sum,
\begin{align*}
 h(\rv{Y}_{1,i}) &\leq \frac{1}{2} \log (2\pi e) \Exp{\rv{Y}_{1,i}^2} \\
 &= \frac{1}{2} \log (2\pi e) \left(\Exp{\rv{X}_{1,i}^2}+\eta_1^2 \Exp{\rv{X}_{0,i}^2}+2 \Exp{\rv{X}_{0,i}\rv{S}_{1,i}}+Q_1+1\right).
\end{align*}
Hence,
\begin{align*}
\sum_{i=1}^{n} h(\rv{Y}_{1,i}) &\leq n \frac{1}{n}\sum_{i=1}^{n} \frac{1}{2} \log (2\pi e) \left(\Exp{\rv{X}_{1,i}^2}+\eta_1^2 \Exp{\rv{X}_{0,i}^2}+2 \eta_1 \Exp{\rv{X}_{0,i}\rv{S}_{1,i}}+Q_1+1\right) \\
&\leq n \frac{1}{2} \log (2\pi e) \left(\frac{1}{n}\sum_{i=1}^{n}\Exp{\rv{X}_{1,i}^2}+\eta_1^2 \frac{1}{n}\sum_{i=1}^{n}\Exp{\rv{X}_{0,i}^2}+2 \eta_1 \frac{1}{n}\sum_{i=1}^{n} \Exp{\rv{X}_{0,i}\rv{S}_{1,i}}+Q_1+1\right) \\
&\leq n \frac{1}{2} \log (2\pi e) \left(P_1+\eta_1^2 P_0+2 \eta_1 \rho_{0S_1} \sqrt{P_0 Q_1}+Q_1+1\right),
\end{align*}
and
\begin{equation*}
nR_1+I(\bv{S}^n;\rv{Y}_1^n|\rv{M}_1) \leq \frac{n}{2} \log \left( P_1+\eta_1^2 P_0+2 \eta_1 \rho_{0S_1} \sqrt{P_0 Q_1}+Q_1+1 \right)+n \epsilon_n
\end{equation*}
$ I(\bv{S}^n;\rv{Y}_1^n|\rv{M}_1)  $ can be written as follows:
\begin{align*}
I(\bv{S}^n;\rv{Y}_1^n|\rv{M}_1) &= h(\bv{S}^n|\rv{M}_1)-h(\bv{S}^n|\rv{M}_1,\rv{Y}_1^n)  \\
&\eqann[=]{a} h(\bv{S}^n)-h(\bv{S}^n|\tilde{\rv{Y}}_1^n) \\
&\eqann[\geq]{b} n h(\bv{S})-\sum_{i=1}^{n} h(\bv{S}_i|\tilde{\rv{Y}}_{1,i}) 
\end{align*}
where \eqannref{a} holds since $ \rv{S}^n $ is independent of $ \rv{M}_1 $ and $ \tilde{\rv{Y}}_1^n \triangleq \eta_1 \rv{X}_0^n+\rv{S}_1^n+\rv{Z}_1^n $, and \eqannref{b} is due to the memorylessness of the state source $ P_{\rv{S}} $ and owing to the fact that conditioning reduces differential entropy . Consider the conditional differential entropy term in the last sum,
\begin{align*}
h(\bv{S}_i|\tilde{\rv{Y}}_{1,i}) &= h\left(\bv{S}_i-\begin{pmatrix}
\kappa_{11} \\ \kappa_{12}
\end{pmatrix} \tilde{\rv{Y}}_{1,i}|\tilde{\rv{Y}}_{1,i} \right)\\
&\leq h\left(\bv{S}_i-\begin{pmatrix}
\kappa_{11} \\ \kappa_{12}
\end{pmatrix} \tilde{\rv{Y}}_{1,i} \right)\\
&\leq \frac{1}{2} \log (2\pi e)^2
\begin{pmatrix}
(1-\kappa_{11})^2 Q_1 Q_2-\eta_1^2 (\kappa_{11} \Exp{\rv{X}_{0,i}\rv{S}_{2,i}}-\kappa_{12} \Exp{\rv{X}_{0,i}\rv{S}_{1,i}}) \\
+ \kappa_{11} Q_2(\kappa_{11}-2\eta_1 \Exp{\rv{X}_{0,i}\rv{S}_{1,i}}(1-\kappa_{11})+\kappa_{11}\eta_1^2 \Exp{\rv{X}_{0,i}^2}) \\
+ \kappa_{12} Q_1(\kappa_{12}-2\eta_1 \Exp{\rv{X}_{0,i}\rv{S}_{2,i}}(1-\kappa_{11})+\kappa_{12}\eta_1^2 \Exp{\rv{X}_{0,i}^2})
\end{pmatrix}
\end{align*}
for any $ \kappa_{11},\kappa_{12} \in \mathbb{R} $. Thus,
\begin{align*}
n \frac{1}{n}\sum_{i=1}^{n} h(\bv{S}_i|\tilde{\rv{Y}}_{1,i}) &\leq \frac{n}{2} \frac{1}{n}\sum_{i=1}^{n}  \log (2\pi e)^2
\begin{pmatrix}
(1-\kappa_{11})^2 Q_1 Q_2-\eta_1^2 (\kappa_{11} \Exp{\rv{X}_{0,i}\rv{S}_{2,i}}-\kappa_{12} \Exp{\rv{X}_{0,i}\rv{S}_{1,i}}) \\
+ \kappa_{11} Q_2(\kappa_{11}-2\eta_1 \Exp{\rv{X}_{0,i}\rv{S}_{1,i}}(1-\kappa_{11})+\kappa_{11}\eta_1^2 \Exp{\rv{X}_{0,i}^2}) \\
+ \kappa_{12} Q_1(\kappa_{12}-2\eta_1 \Exp{\rv{X}_{0,i}\rv{S}_{2,i}}(1-\kappa_{11})+\kappa_{12}\eta_1^2 \Exp{\rv{X}_{0,i}^2})
\end{pmatrix} \\
&\leq \frac{n}{2} \log (2\pi e)^2
\begin{pmatrix}
(1-\kappa_{11})^2 Q_1 Q_2-\eta_1^2 (\kappa_{11} \frac{1}{n}\sum_{i=1}^{n} \Exp{\rv{X}_{0,i}\rv{S}_{2,i}}-\kappa_{12} \frac{1}{n}\sum_{i=1}^{n} \Exp{\rv{X}_{0,i}\rv{S}_{1,i}}) \\
+ \kappa_{11} Q_2(\kappa_{11}-2\eta_1 \frac{1}{n}\sum_{i=1}^{n} \Exp{\rv{X}_{0,i}\rv{S}_{1,i}}(1-\kappa_{11})+\kappa_{11}\eta_1^2 \frac{1}{n}\sum_{i=1}^{n} \Exp{\rv{X}_{0,i}^2}) \\
+ \kappa_{12} Q_1(\kappa_{12}-2\eta_1 \frac{1}{n}\sum_{i=1}^{n} \Exp{\rv{X}_{0,i}\rv{S}_{2,i}}(1-\kappa_{11})+\kappa_{12}\eta_1^2 \frac{1}{n}\sum_{i=1}^{n} \Exp{\rv{X}_{0,i}^2})
\end{pmatrix} \\
&\leq \frac{n}{2} \log (2\pi e)^2
\begin{pmatrix}
(1-\kappa_{11})^2 Q_1 Q_2-\eta_1^2 (\kappa_{11} \rho_{0S_2} \sqrt{P_0Q_2}-\kappa_{12} \rho_{0S_1} \sqrt{P_0Q_1}) \\
+ \kappa_{11} Q_2(\kappa_{11}-2\eta_1 \rho_{0S_1} \sqrt{P_0Q_1}(1-\kappa_{11})+\kappa_{11}\eta_1^2 P_0 \\
+ \kappa_{12} Q_1(\kappa_{12}-2\eta_1 \rho_{0S_2} \sqrt{P_0Q_2}(1-\kappa_{11})+\kappa_{12}\eta_1^2 P_0)
\end{pmatrix}.
\end{align*}
Next, we choose $ \kappa_{11} $ and $ \kappa_{12} $ as follows:
\begin{align*}
\kappa_{11} &= \frac{\eta_1 \rho_{0S_1} \sqrt{P_0Q_1}+Q_1}{\eta_1^2P_0+2\eta_1 \rho_{0S_1} \sqrt{P_0Q_1}+Q_1+1} \\
\kappa_{12} &= \frac{\eta_1 \rho_{0S_2} \sqrt{P_0Q_2}}{\eta_1^2P_0+2\eta_1 \rho_{0S_1} \sqrt{P_0Q_1}+Q_1+1}
\end{align*}
and evaluate the latter inequality to have
\begin{align*}
n \frac{1}{n}\sum_{i=1}^{n} h(\bv{S}_i|\tilde{\rv{Y}}_{1,i}) &\leq \frac{n}{2} \log (2\pi e)^2 \left( \frac{Q_1Q_2 \left((1-\rho_{0S_1}^2-\rho_{0S_2}^2)\eta_1^2 P_0+1\right)}{\eta_1^2P_0+2\eta_1\rho_{0S_1}\sqrt{P_0Q_1}+Q_1+1}\right)
\end{align*}
Finally
\begin{align*}
I(\bv{S}^n;\rv{Y}_1^n|\rv{M}_1) \geq  \frac{n}{2} \log  \frac{\eta_1^2 P_0+2\rho_{0s_1} \eta_1 \sqrt{P_0 Q_1}+Q_1+1}{(1-\rho_{0s_1}^2-\rho_{0s_2}^2) \eta_1^2 P_0+1}
\end{align*}
Thus the bound in  \eqref{eq:independent_states_R1_upper_bound} is satisfied. The bound \eqref{eq:independent_states_R2_upper_bound} follows from similar considerations.
It remains to show that $ \rho_{0S_1}^2+\rho_{0S_2}^2 \leq 1 $. We use the non-negativity property of the covariance matrix of $ (\rv{X}_0,\rv{S}_1,\rv{S}_2) $
\begin{align*}
\det  \Sigma_{\rv{X}_0 \rv{S}_1 \rv{S}_2}  &= \det \begin{bmatrix}
\var{\rv{X}_0} & \cov{\rv{X}_0 \rv{S}_1} & \cov{\rv{X}_0 \rv{S}_2} \\
\cov{\rv{X}_0 \rv{S}_1} & \var{\rv{S}_1} & \cov{\rv{S}_1 \rv{S}_2} \\
\cov{\rv{X}_0 \rv{S}_2} & \cov{\rv{S}_1 \rv{S}_2} & \var{ \rv{S}_2} \\
\end{bmatrix} \\
& \eqann{a} \det \begin{bmatrix}
\var{\rv{X}_0} & \cov{\rv{X}_0 \rv{S}_1} & \cov{\rv{X}_0 \rv{S}_2} \\
\cov{\rv{X}_0 \rv{S}_1} & \var{\rv{S}_1} & 0 \\
\cov{\rv{X}_0 \rv{S}_2} & 0 & \var{ \rv{S}_2} \\
\end{bmatrix} \\
&= \var{\rv{X}_0} \var{\rv{S}_1} \var{\rv{S}_2}\\
&\ - (\cov{\rv{X}_0 \rv{S}_1})^2 \var{\rv{S}_2}- (\cov{\rv{X}_0 \rv{S}_2})^2 \var{\rv{S}_1}\\
& \eqann[\geq]{b} 0
\end{align*}
where \eqannref{a} follows from independence of $ \rv{S}_1 $ and $ \rv{S}_2 $ and $ \eqannref{b} $ from non-negativity of covariance matrix. Now we arrange parts and use \eqref{eq:correlation_coefficients_def} to have:
\begin{align*}
\frac{(\cov{\rv{X}_0 \rv{S}_1})^2}{\var{\rv{X}_0} \var{\rv{S}_1}}+\frac{(\cov{\rv{X}_0 \rv{S}_2})^2}{\var{\rv{X}_0} \var{\rv{S}_2}} = \rho_{0S_1}^2+\rho_{0S_2}^2 \leq 1.
\end{align*}
This completes the proof of Proposition \ref{proposition:independent_states_outer_bound}.

\section{} \label{apndx:independent_state_inner_bound}
We use random codes and fix the following joint distribution:
\begin{align*}
P_{SUVX_0 X_1 X_2 Y_1 Y_2} = P_{SUV} P_{X_0|SUV} P_{X_1} P_{X_2} P_{Y_1|S X_0 X_1} P_{Y_2|S X_0 X_2}
\end{align*}
\textbf{Codebook Generation.} Generate $ 2^{n\tilde{R}_U} $  randomly and independently generated sequences $ u^n(r) $, $ r\in[1:2^{n\tilde{R}_U}] $, each according to $ \prod_{i=1}^n P_U (u_i) $. Similarly, generate $ 2^{n\tilde{R}_{V}} $ randomly and independently generated sequences $ v^n(t) $, $ t \in [1:2^{n\tilde{R}_{V}}] $ according to $ \prod_{i=1}^n P_{V} (v_i) $. 

Randomly and independently generate $ 2^{nR_1} $ sequences $ x_1^n(m_1) $, $ m_1 \in [1:2^{nR_1}] $, each according to $ \prod_{i=1}^n P_{X_1} (x_{1i}) $. Similarly generate $ 2^{nR_2} $ sequences $ x_2^n(m_2) $, $ m_2 \in [1:2^{nR_2}] $, each according to $ \prod_{i=1}^n P_{X_2} (x_{2i}) $.

These codewords constitute the codebook, which is revealed to the encoders and the decoders.

\textbf{Encoding.}
Let $ \epsilon''>\epsilon'>0 $.The encoder at the helper, given $ s^n $, finds $ \tilde{r} $ such that 
\[ (s^n,u^n(\tilde{r})) \in \mathcal{T}_{\epsilon'}^{(n)} (P_{SU}), \] if there is more then one such $ \tilde{r} $, choose the smallest one. If no such $ \tilde{r} $ can be found declare an error. Next, given $ (s^n,u^n(\tilde{r})) $, find $ \tilde{t} $ such that
\[ \big(s^n, u^n(\tilde{r}),v^n(\tilde{t}) \big) \in \mathcal{T}_{\epsilon''}^{(n)} (P_{SUV}) \]
if there is more then one such $ \tilde{t} $, choose the smallest one. If no such $ \tilde{t} $ can be found declare an error. Then, given $ s^n $, $ u^n(\tilde{r}) $ and $ v^n(\tilde{t}) $, generate $ x_0^n $ with i.i.d. components according to $ \prod_{i=1}^{n} P_{\rv{X}_0|\rv{S}\rv{U} \rv{V}} (x_{0,i}|s_i,u_i,v_i) $.

To send message $ m_1 $, the encoder at transmitter 1 transmits $ x_1^n(m_1) $. Similarly, to send message $ m_2 $, the encoder at transmitter 2 transmits $ x_2^n(m_2) $.

\textbf{Decoding.}
Let $ \epsilon>\epsilon''>\epsilon'>0 $. Given $ y_1^n $, decoder 1 declares that $ \hat{m}_1 $ was sent if its the unique message such that 
\[ (u^n(\hat{r}),x_1^n(\hat{m}_1),y_1^n) \in \mathcal{T}_{\epsilon}^{(n)} (P_{UX_1Y_1}). \] If no or more than one $ \hat{m}_1 $ can be found, it declares an error. 

Similarly, given  $ y_2^n $, decoder 2 finds the unique message  $ \hat{m}_2 $ such that 
\[ ( v^n(\hat{t}),x_2^n(\hat{m}_2),y_2^n) \in \mathcal{T}_{\epsilon}^{(n)} (P_{V X_2 Y_2}). \] If no or more than one $ \hat{m}_2 $ can be found, it declares an error.

\textbf{Analysis of the probability of error.} Assume without loss of generality that the message pair $ (\rv{M}_1,\rv{M}_2) = (1,1) $ was sent and let $ r_0 $ be the chosen index for $ u^n $ and $ t_0 $ be the chosen index for $ v^n $. The encoder at the helper makes an error only if one or both of the following errors occur:
\begin{align*}
\mathcal{E}_{01} &= \{ (\rv{S}^n,\rv{U}^n(r))\notin \mathcal{T}_{\epsilon'}^{(n)}(P_{\rv{S}\rv{U}}) \text{ for all } r\in [1:2^{n\tilde{R}_U}] \}, \\
\mathcal{E}_{02} &= \{ (\rv{S}^n,\rv{U}^n(r_0),\rv{V}^n(t))\notin \mathcal{T}_{\epsilon'}^{(n)}(P_{\rv{S}\rv{U}\rv{V}}) \text{ for all } t\in [1:2^{n\tilde{R}_V}] \}.
\end{align*}
Thus, by the union of events bound, the probability that the encoder at the helper makes an error, can be upper bounded as
\begin{equation*}
\Pr(\mathcal{E}_0) = \Pr (\mathcal{E}_{01} \cup \mathcal{E}_{02}) \leq \Pr (\mathcal{E}_{01})+\Pr(\mathcal{E}_{01}^c \cap \mathcal{E}_{02})
\end{equation*}
By the covering lemma with $ \rv{U} = \emptyset $, $ \rv{X} \leftarrow \rv{S} $,  $ \hat{\rv{X}}\leftarrow \rv{U} $, and $ \mathcal{A}=\{1:2^{n\tilde{R}_U}\} $, $ \Pr(\mathcal{E}_{01}) $ tends to zero as $ n\rightarrow \infty $ if $ \tilde{R}_U > I(\rv{U};\rv{S})+\delta(\epsilon') $.

Similarly, using the covering lemma with $ \rv{U} = \emptyset $, $ \rv{X} \leftarrow (\rv{S},\rv{U}) $, $ \hat{\rv{X}} \leftarrow \rv{V} $, and $ \mathcal{A} = \{1:2^{n\tilde{R}_V}\} $, $ \Pr(\mathcal{E}_{01}^c \cap \mathcal{E}_{02}) $ tends to zero as $ n\rightarrow \infty $ if $ \tilde{R}_V > I(\rv{V};\rv{S},\rv{U}) + \delta(\epsilon'') $.

The decoder at receiver 1 makes an error only if one or more of the following events occur
\begin{align*}
\mathcal{E}_{11} &= \{ (\rv{U}^n(r_0),\rv{X}_1^n(1),\rv{Y}_1^n)\notin \mathcal{T}_{\epsilon}^{(n)}(P_{\rv{U}\rv{X}_1\rv{Y}_1})  \}, \\
\mathcal{E}_{12} &= \{ (\rv{U}^n(r_0),\rv{X}_1^n(m_1),\rv{Y}_1^n)\in \mathcal{T}_{\epsilon}^{(n)}(P_{\rv{U}\rv{X}_1\rv{Y}_1}) \text{ for some } m_1 \neq 1 \},\\
\mathcal{E}_{13} &= \{ (\rv{U}^n(r),\rv{X}_1^n(m_1),\rv{Y}_1^n)\in \mathcal{T}_{\epsilon}^{(n)}(P_{\rv{U}\rv{X}_1\rv{Y}_1}) \text{ for some } r\neq r_0 \text{ and } m_1 \neq 1 \}.
\end{align*}
Again, by the union of events bound, the probability that the decoder at receiver 1 makes an error, can be upper bounded as
\begin{align*}
\Pr(\mathcal{E}_1) &= \Pr ( \mathcal{E}_{11} \cup \mathcal{E}_{12}\cup \mathcal{E}_{13}) \\
&\leq  \Pr (\mathcal{E}_{01} \cup \mathcal{E}_{11} \cup \mathcal{E}_{12}\cup \mathcal{E}_{13}) \\
&\leq \Pr (\mathcal{E}_{01})+\Pr(\mathcal{E}_{01}^c \cap \mathcal{E}_{11})+\Pr(\mathcal{E}_{01}^c \cap \mathcal{E}_{12})+\Pr(\mathcal{E}_{13})
\end{align*}
We have already shown that $ \Pr (\mathcal{E}_{01}) $ tends to zero as $ n\rightarrow \infty $ if $ \tilde{R}_U > I(\rv{U};\rv{S})+\delta(\epsilon') $. Next, note that
\begin{equation*}
\mathcal{E}_{01}^c = \{ (\rv{S}^n,\rv{U}^n(r_0))\in \mathcal{T}_{\epsilon'}^{(n)}(P_{\rv{S}\rv{U}})  \} = \{ (\rv{S}^n,\rv{U}^n(r_0),\rv{X}_0^n)\in \mathcal{T}_{\epsilon'}^{(n)}(P_{\rv{S}\rv{U}\rv{X}_0})  \}
\end{equation*}
and
\begin{align*}
P_{\rv{Y}_1^n|\rv{S}^n \rv{U}^n(r_0) \rv{X}_0^n\rv{X}_1^n(1)} (y_1^n|s^n,u^n,x_0^n.x_1^n) &= \prod_{i=1}^n P_{\rv{Y}_1|\rv{S}\rv{U}\rv{X}_0\rv{X}_1} (y_{1i}|s_i,u_i,x_{0i},x_{1i}) \\
&= \prod_{i=1}^n P_{\rv{Y}_1|\rv{S}\rv{X}_0\rv{X}_1} (y_{1i}|s_i,x_{0i},x_{1i})
\end{align*}
Hence, by the conditionally typicality lemma, $ \Pr(\mathcal{E}_{01}^c \cap \mathcal{E}_{11}) $ tends to zero as $ n\rightarrow \infty $.

As for the probability of the event $ (\mathcal{E}_{01}^c \cap \mathcal{E}_{12}) $, $ \rv{X}_1^n(m_1) $ is independent of $ (U^n(r0),\rv{Y}_1^n) \sim \prod_{i=1}^{n} P_{\rv{U}\rv{Y}_1} (u_i,y_{1i})$. Hence, by the packing lemma, with $ \rv{U} = \emptyset $, $ \rv{X} \leftarrow (\rv{U},\rv{X}_1) $, $ \rv{Y} \leftarrow \rv{Y}_1^n $ and $ \mathcal{A} = \{[1:2^{n\tilde{R}_U}]/r_0 \times [2:2^{nR_1}] \} $, $ \Pr (\mathcal{E}_{01}^c \cap \mathcal{E}_{12})  $ tends to zero as $ n\rightarrow $ if $ R_1 < I(\rv{X}_1;\rv{U},\rv{Y}_1) -\delta(\epsilon) $. $ \rv{X}_1 $ and $ \rv{U} $ are mutually independent, hence the latter condition is equivalent to $ R_1 < I(\rv{X}_1;\rv{Y}_1|\rv{U}) -\delta(\epsilon) $.

Finally, since for $ m_1 \neq 1 $, $ r \neq r_0 $, $ (\rv{X}_1^n(m_1),\rv{U}^n(r)) $ is independent of $ (\rv{X}_1^n(1),\rv{U}^n(r_0),\rv{Y}_1^n) $, again by the packing lemma with $ \rv{U} = \emptyset $, $ \rv{X} \leftarrow (\rv{U},\rv{X}_1) $, $ \rv{Y} \leftarrow \rv{Y}_1 $ and $ \mathcal{A}= [2:2^{nR_1}] \times [1:2^{n\tilde{R}_U}]/r_0 $, $ \Pr (\mathcal{E}_{13}) $ tends to zero as $ n\rightarrow \infty $ if $ \tilde{R}_U+R_1 < I(\rv{U},\rv{X}_1;\rv{Y}_1) -\delta(\epsilon) $.

Next consider the average probability of error for decoder 2. The decoder at receiver 2 makes an error only if one or more of the following events occur
\begin{align*}
\mathcal{E}_{21} &= \{ (\rv{V}^n(t_0),\rv{X}_2^n(1),\rv{Y}_2^n)\notin \mathcal{T}_{\epsilon}^{(n)}(P_{\rv{V}\rv{X}_2\rv{Y}_2})  \}, \\
\mathcal{E}_{22} &= \{ (\rv{V}^n(t_0),\rv{X}_2^n(m_2),\rv{Y}_2^n)\in \mathcal{T}_{\epsilon}^{(n)}(P_{\rv{V}\rv{X}_2\rv{Y}_2}) \text{ for some } m_2 \neq 1 \},\\
\mathcal{E}_{23} &= \{ (\rv{V}^n(t),\rv{X}_2^n(m_2),\rv{Y}_2^n)\in \mathcal{T}_{\epsilon}^{(n)}(P_{\rv{V}\rv{X}_2\rv{Y}_2}) \text{ for some } t\neq t_0 \text{ and } m_2 \neq 1 \}.
\end{align*}
Again, by the union of events bound, the probability that the decoder at receiver 2 makes an error, can be upper bounded as
\begin{align*}
\Pr(\mathcal{E}_2) &= \Pr ( \mathcal{E}_{21} \cup \mathcal{E}_{22}\cup \mathcal{E}_{23}) \\
&\leq  \Pr (\mathcal{E}_{0} \cup \mathcal{E}_{21} \cup \mathcal{E}_{22}\cup \mathcal{E}_{23}) \\
&\leq \Pr (\mathcal{E}_{0})+\Pr(\mathcal{E}_{0}^c \cap \mathcal{E}_{21})+\Pr(\mathcal{E}_{0}^c \cap \mathcal{E}_{22})+\Pr(\mathcal{E}_{23})
\end{align*}
In a very similar fashion as was shown for decoder 1, it can be shown that $ \Pr(\mathcal{E}_2) $ tends to zero as $ n\rightarrow \infty $ if
\begin{align*}
\tilde{R}_V \geq&\ I(\rv{V};\rv{S},\rv{U})+\delta(\epsilon') \\
R_2 \leq&\ I(\rv{X}_2;\rv{Y}_2|\rv{V}) -\delta(\epsilon)\\
R_2+\tilde{R}_{V} \leq&\ I(\rv{V},\rv{X}_2;\rv{Y}_2) - \delta(\epsilon)
\end{align*}
Finally, combining the aforementioned bounds  yields the following achievable region:
\begin{align*}
R_1 \leq&\ \min \big\{I(\rv{U},\rv{X}_1;\rv{Y}_1)- I(\rv{U};\rv{S}),I(\rv{X}_1;\rv{Y}_1|\rv{U}) \big\} \\
R_2 \leq&\ \min \big\{I(\rv{V},\rv{X}_2;\rv{Y}_2)- I(\rv{V};\rv{U},\rv{S}),I(\rv{X}_2;\rv{Y}_2|\rv{V})\big\}
\end{align*}
This completes the proof of achievability.

\bibliography{bibliography}
\bibliographystyle{unsrt}
%\nocite{Costa83}
%\vspace{-0.15cm}

% trigger a \newpage just before the given reference
% number - used to balance the columns on the last page
% adjust value as needed - may need to be readjusted if
% the document is modified later
%\IEEEtriggeratref{8}
% The "triggered" command can be changed if desired:
%\IEEEtriggercmd{\enlargethispage{-5cm}}

\end{document}